\documentclass[]{article}

\def\vec#1{{\mathbf#1}}

\begin{document}

\title{Asymptotic Zero Energy States for SU(N $\geq$ 3)}

\author{Jens Hoppe\\
\small\it Albert Einstein Institute\\
\small\it  Am M\"uhlenberg 1\\
\small\it  D~14476 Golm}
\date{}
\maketitle

{\narrower\abstract{\noindent Some ideas are presented
concerning the question which of the harmonic wavefunctions
constructed in [hep-th/9909191] may be annihilated by all
supercharges.}\par}

\bigskip

In an attempt to extend our knowledge about the asymptotic form
of zero-energy wave functions of SU(N) invariant supersymmetric
matrix models beyond $N=2$, it was recently shown \cite{1}, for
$N=3$, how to construct Weyl~$\otimes$~Spin($d$) invariant
asymptotic states out of the Cartan-subalgebra degrees of freedom.
To find out which of these harmonic wavefunctions is annihilated
by the asymptotic supercharges~\cite{2}
\begin{equation} Q_\beta = -i
\gamma^t_{\beta\alpha}\nabla_{tk}\Theta_{\alpha k}
\end{equation}
\begin{eqnarray*} t & = & 1, \ldots, d=(2),3,5 \mbox{ or } 9 \\
\alpha,\beta & = & 1,\ldots, s_d = 2,4,8 \mbox{ resp. } 16 \\ k &=&
1,2
\end{eqnarray*} is non-trivial. The ``guess'' presented in this note
will hopefully be a first step\footnote{A detailed calculation of the
matrix element (13) is under investigation by J.~Plefka.}. Should
the answer really be that for arbitrary $N$, already for this  ``free"
problem, exactly one supersymmetric state exists for $d=9$ (and
none for
$d=2,3,5$), this should obviously have a ``simple" (deep?)\ 
mathematical explanation, of more general relevance.

Each harmonic state, constructed in \cite{1}, has the form
\begin{equation}
\Psi = \sum_{l,S,R,m}r^{-2l-2(d-1)}\bar\Psi_{lm}^{S\times
R}(\vec{x}_1,\vec{x}_2) | S\times R;m\rangle
\end{equation} with $l$=0,1,\ldots,$\bar\Psi_{lm}^{S\times R}$ a
harmonic polynomial of degree $l$ (in the variables
$\vec{x}_1,\vec{x}_2$; $r:=\sqrt{\vec{x}_1^2 +
\vec{x}_2^2})$ transforming under the Weyl-group (the
permutation group
$S_3$) and Spin($d$) according to the irreducible representation
$ S (=1, \in$ or $ \rho)$, cf.~\cite{1},   resp.\ $R,  m = 1,2,\ldots
\dim(S\times R), \mbox{ and }| S\times R;m\rangle 
$ is the corresponding basis-vector in a $S\times R$ representation
present in the fermionic Fock-space ${\mathcal H}(d)={\mathcal
H}_{2^{\frac{1}{2}S_D}}\times {\mathcal H}_{2^{\frac{1}{2}S_D}}$
(cf.~\cite{1}). As pointed out by M. Bordemann, one way to
construct a state that will be annihilated by all the supercharges, is
to let
\begin{equation}
\Psi = \left( \prod^{s_d}_{\beta =1} Q_\beta \right) \Phi
\end{equation} with $\Phi$ being any of the harmonic states (2).
The crucial question is: for which $\Phi$ will (3) be non-zero? Let
$d=9$ now ($s_d =16)$. The guess which I would like to discuss
here, is that
\begin{equation}
\Phi = r^{-16} | 1\rangle
\end{equation} will do, where $|1\rangle$ is the unique
Weyl~$\times$~Spin(9) invariant state in
${\mathcal H}= {\mathcal H}_{256}\otimes {\mathcal H}_{256}.$
Why (4)? One simple reason(ing) is the following: As each
$Q_\beta$, acting on the product of a harmonic, homogenous
polynomial and a negative power of
$r$ will {\em increase} the degree of the polynomial, (4) is the only
harmonic state which certainly (a priori!)\ can \emph{not} be the
image of
$Q_\beta$ acting on some harmonic
$\chi$. Actually, if we could show that all harmonic $\Phi$'s not
containing a contribution from $l=0$ are of the form 
\begin{equation}
\Phi=\sum_{\rho}Q_\beta \Phi_\beta
\end{equation} with $\Phi_\beta$ harmonic, (4) would necessarily
be the only chance left, as (3) is clearly identically zero, if $\Phi$ is
of the form~(5).

In any case, consider now
\begin{equation}
\Psi:=\epsilon_{\beta_1\cdots\beta_{16}}Q_{\beta_1}\cdot
Q_{\beta_2}\cdots 
 Q_{\beta_{16}}\frac{1}{r^{16}}|1\rangle
\end{equation} Is it zero? First of all, one needs to know more
explicitly, what the state $|1\rangle \in {\mathcal H}$ is.

As $ {\mathcal H}_{256}$ contains only 3 irreducible Spin(9)
representations, 
$$ {\mathcal H}_{256} = 44 \oplus 84 \oplus 128
$$
${\mathcal H}$ contains only 3 Spin(9) singlets, namely
\begin{eqnarray}\nonumber |1\rangle_{44} &:=&
\sum_{s,t}|st\rangle |st\rangle' \\ |1\rangle_{84} &:=&
\sum_{s,t,u}|stu\rangle |stu\rangle' \\ \nonumber |1\rangle_{128}
&:=& \sum_{t,\alpha}|t\alpha\rangle |t\alpha\rangle'\ . 
\end{eqnarray} For notational simplicity, the fermions
$\Theta_{\alpha k = 2}$ are sometimes denoted by
$\Theta'_\alpha$, and $|st\rangle = |ts\rangle$ 
$(\sum_s |ss\rangle =0$), $|stu\rangle$ (totally antisymmetric in
$s,t,u$) and
$|t\alpha\rangle
$(with $\gamma^t_{\beta\alpha}|t\alpha\rangle=0)$ stands for the
basis-elements of the 44,84, resp. 128-dimensional representation.

Defining fermionic creation operators
\begin{equation}
\lambda_{\alpha k}:= \frac{1}{\sqrt{ 2}}(\Theta_{\alpha k} + i
\Theta_{\alpha +8,k})_{\alpha=1,\ldots,8}
\end{equation} together with the representation
$$
\gamma^9 = \left(\begin{array}{cc} \mathbf{1} & 0 \\ 0&
-\mathbf{1} 
\end{array}\right),\quad
\gamma^8 = \left(\begin{array}{cc} 0 &\mathbf{1}  \\ \mathbf{1}
& 0 
\end{array}\right),\quad
\gamma^j = \left(\begin{array}{cc} 0& i\Gamma^j \\ - i\Gamma^j &
0
\end{array}\right),\quad
$$
$$ (i\Gamma^j)_{k8}:=\delta_{jk}, \quad (i\Gamma^j)_{kl}:=-c_{jkl},
$$ and totally antisymmetric `octonionic structure constants'
$c_{jkl} = +1$  for $(ijk)=123$, 147, 165, 246, 257, 354, 367,  the 3
states in (7) may also be explicitely given as concrete polynomials
in the creation operators $\lambda_{\alpha k}$. E.g., with
\begin{equation}
b_j:=\frac{i}{4}\lambda_\alpha\Gamma^i_{\alpha\beta}
\lambda_{\beta},
\quad
c_j:=\frac{i}{4}\lambda'_\alpha\Gamma^i_{\alpha\beta}\lambda'_\beta
\end{equation} one finds
\begin{equation}
|1\rangle_{44}=\Bigl((\vec{b}\cdot\vec{c})^2-\frac{1}{9}\vec{b}^2\vec{c}^2-
\frac{2}{9}\vec{b}\cdot\vec{c}(b^2+c^2)+\frac{2}{63}(b^4+c^4)\Bigr)|0\rangle
\end{equation} while the states $|st\rangle$ are explicitly given as
follows
$(|8\rangle := \lambda_1 \cdots \lambda_8|0\rangle)$
\begin{eqnarray}\nonumber  |i\neq j\rangle &=& b_ib_j |0\rangle
\\\nonumber  |jj\rangle &=& (b_j^2 - \frac{1}{9} \vec{b}^2)|0\rangle
\\\nonumber  |j9\rangle &=& -\frac{i}{2}
(b_j+\frac{2}{9}b_j\vec{b}^2)|0\rangle \\\nonumber  |j8\rangle 
&=& \frac{1}{2} (b_j-\frac{2}{9}b_j\vec{b}^2)|0\rangle \\
|89\rangle &=& -\frac{i}{2} (|0\rangle - |8\rangle) \\ |88\rangle
&=& -\frac{1}{2} (-|0\rangle + \frac{2}{9}\vec{b}^2|0\rangle -
|8\rangle) \\\nonumber  |99\rangle &=& -\frac{1}{2}  (|0\rangle +
\frac{2}{9}\vec{b}^2|0\rangle + |8\rangle)\nonumber 
\end{eqnarray} In any case, as one of the Weyl-transformations
changes
$\lambda_\alpha'$ to -$\lambda_\alpha'$ (while leaving
$\lambda_\alpha$ invariant), $|1\rangle_{128}$ can not be
contained in the Weyl-invariant state  $|1\rangle$, which therefore
must be a linear combination of $|1\rangle_{44}$ and
$|1\rangle_{84}$

Projecting (6) onto this linear combination will give some
Weyl~$\times$~Spin(9) invariant differential operator of degree 16
(with constant coefficients), acting on $r^{-16}$. While R.~Suter and I
checked, by using quite different methods, that a priori only 2 such
independent operators, not containing the full Laplace-operator
(which annihilates $r^{-16}!$) exist, one needs to know
\begin{equation}
\langle1|\Theta_{\alpha_1 k_1}\Theta_{\alpha_2 k_2} \cdots
\Theta_{\alpha_{16}k_{16}} |1\rangle
\end{equation} resp.~the contraction with $\epsilon_{\beta_1\cdots
\beta_{16}}\gamma^{t_1}_{\beta_1 \alpha_1}\cdots
\gamma^{t_{16}}_{\beta_{16}\alpha_{16}}$ (times
$\nabla_{t_1k_1}\cdots\nabla_{t_{16}k_{16}}r^{-16}$).

Should the result turn out to be non-zero, (6) will, by construction,
be a non-trivial supersymmetric wave function. For general $N>2$
the corresponding asymptotic fall off would be $r^{-((N-1)d+14)}$. 

A simpler way to describe the fermionic part of the wavefunction
is to define fermionic creation operators
$$
\Lambda_\alpha = \frac1{\sqrt2} (\theta_{\alpha_1} + i
\theta_{\alpha_2}) , \quad 
\alpha=1,\ldots,16,
$$ and to observe that 
$$
\gamma^{uv}_{\alpha_1\alpha_2}
\gamma^{vp}_{\alpha_3\alpha_4}
\gamma^{pq}_{\alpha_5\alpha_6}
\gamma^{qu}_{\alpha_7\alpha_8} 
\Lambda_{\alpha_1}\Lambda_{\alpha_2}\cdots\Lambda_{\alpha_8}
|0\rangle
$$ is Spin(9)~$\times$~Weyl invariant.

\subsubsection*{Acknowledgement}

I would like to thank M. Bordemann, J. Plefka, A. Smilga, and R.
Suter for valuable discussions, as well as ETH Z\"urich, the Korean
Institute for Advanced Studies,  Brown University, the MIT Center
for Theoretical Physics,  and Harvard University for kind
hospitality.


\begin{thebibliography}{99}

\bibitem{1} M. Bordemann, J. Hoppe, R. Suter; hep-th/9909191

\bibitem{2} V.Kac, A. Smilga; hep-th/9908096.

\end{thebibliography}
\end{document}